\documentclass[12pt,thmsa,arabtex]{article}
\usepackage{amsfonts}
\usepackage{amsmath}
\usepackage{graphicx}

\input{tcilatex}

\begin{document}

\begin{center}
{\Large Sudden death and long-living of entanglement in an
ion-trap laser }

M. Abdel-Aty

{\small Department of Mathematics, College of Science, University of
Bahrain, 32038, Kingdom of Bahrain}

\bigskip
\end{center}

The dynamical properties of quantum entanglement in a time-dependent
three-level-trapped ion interacting with a laser field are studied in terms
of the reduced-density linear entropy considering two specific initial
states of the field. Allowing the instantaneous position of the
center-of-mass motion of the ion to be explicitly time-dependent, it is
shown that either sudden death of entanglement or survivability of quantum
entanglement can be obtained with a specific choice of the initial state
parameters. The difference in evolution picture corresponding to the
multi-quanta processes is discussed.

\section{Introduction}

One of the most striking differences between classical and quantum
correlations is the restricted capability of quantum states to share
entanglement. The decay of entanglement cannot be restored by local
operations and classical communications, which is one of the main obstacles
to achieve the quantum computer \cite{ben96}. Therefore it becomes an
important subject to study the loss of entanglement \cite%
{yu06,yon06,pin06,yu02,yu03,yu04}. Quite recently, by using vacuum noise
two-qubit, entanglement is terminated abruptly in a finite time have been
performed \cite{yu06} and the entanglement dynamics of a two two-level atoms
model have been discussed \cite{yon06}. They called the non-smooth
finite-time decay entanglement sudden death.

On the other hand, trapped atomic ions are an ideal system for exploring
quantum information science. Recent advances in the dynamics of trapped ions
(for a recent review, see e.g., \cite{lei03}) have demonstrated that a
macroscopic observer can effectively control dynamics as well as perform a
complete measurement of states of microscopic quantum systems. With the
reliance in the processing of quantum information on a cold trapped ion, a
long-living entanglement in the ion-field interaction with pair cat states
has been observed \cite{abd06} . Also, experimental preparation and
measurement of the the motional state of a trapped ion, which has been
initially laser cooled to the zero-point of motion, has been reported in
\cite{lei97}.

In this paper we present an explicit connection between the initial state
setting of the field and the dynamics of the entanglement. We give a
condition for the existence of either entanglement sudden death or
long-lived entanglement. In particular, a quantitative characterization of a
general system of a three-level trapped ion interacting with a laser field
is presented. We present various numerical examples in order to monitor the
linear entropy and entanglement dynamics. The paper is organized as follows.
In section 2, we consider a general class of \ a three-level system and
obtain its solution. In section 3 we discuss the dynamics of the
entanglement with different initial states. Finally, we summarize the
results and conclude in section 4.

\begin{figure}[tbph]
\begin{center}
\includegraphics[width=13cm]{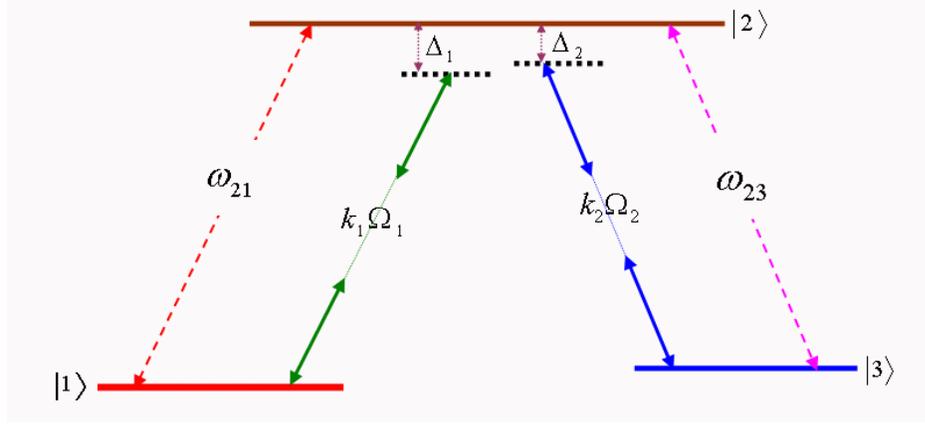}
\end{center}
\caption{Energy-level diagram for a three-state $\Lambda -$type system
interacting with a bimodal cavity field coupling the two ground states $%
|1\rangle $ and $|3\rangle $ to a common excited state $|2\rangle $ via a
Raman transition. }
\end{figure}

\section{Model}

Now let us consider the Hamiltonian which describes a single trapped ion in
a two-dimensional trap. Therefore, the physical system on which we focus is
a three-level harmonically trapped ion with its center-of-mass motion
quantized. We denote by $\hat{\psi}_{i}$ and $\hat{\psi}_{i}^{\dagger }$ the
annihilation and creation operators and $\upsilon _{1}(\upsilon _{2})$ is
the vibrational frequency related to the center-of-mass harmonic motion
along the direction $\hat{x}(\widehat{y}).$ The trapped ion Hamiltonian may
be written as \cite{mes03,yoo85,moy00}
\begin{eqnarray}
\hat{H} &=&\hat{H}_{0}+\hat{H}_{int}(t),  \notag \\
\hat{H}_{0} &=&\hbar \sum\limits_{i=1}^{2}\upsilon _{i}\hat{\psi}%
_{i}^{\dagger }\hat{\psi}_{i}+\hbar \sum\limits_{i=1}^{3}\omega _{i}\hat{S}%
_{ii},  \notag \\
\hat{H}_{int}(t) &=&\hbar \Im _{1}(\hat{x},t)\hat{S}_{12}+\hbar \Im \lambda
_{2}\Im _{2}(\widehat{y},t)\hat{S}_{31}+\hbar \Im _{1}^{\ast }(\hat{x},t)%
\hat{S}_{21}+\hbar \Im _{2}^{\ast }(\widehat{y},t)\hat{S}_{13}.  \label{ham}
\end{eqnarray}%
We denote by $\hat{S}_{lm}$ the atomic flip operator for the $|m\rangle
\rightarrow |l\rangle $ transition between the two electronic states, where $%
\hat{S}_{lm}=|l\rangle \langle m|,(l,m=1,2,3)$.

So far we have disregarded relaxations since we are interested in the
dynamics for short times. Suppose the ion is irradiated by a laser field of
the form
\begin{eqnarray}
\Im _{1}(\hat{x},t) &=&\frac{\epsilon _{1}\langle 1|d_{1}.\wp _{1}|2\rangle
}{\hbar }\exp [-i(k_{1}\hat{x}-\Omega _{1}t)],  \notag \\
\Im _{2}(\hat{y},t) &=&\frac{\epsilon _{2}\langle 1|d_{2}.\wp _{2}|2\rangle
}{\hbar }\exp [-i(k_{1}\hat{y}-\Omega _{2}t)],
\end{eqnarray}%
where $\epsilon _{1}$ and $\epsilon _{2}$ are the amplitudes of the two
laser fields with frequencies $\Omega _{1}$ and $\Omega _{2}$ and
polarization vectors $\wp _{1}$ and $\wp _{2}$, respectively. The transition
in the three-level ion is characterized by the dipole moment $d_{i}$ and $%
k_{i},i=1,2$ are the wave vectors of the two laser fields$.$We define the
detuning between the atomic transitions and the fields as $\Delta
_{1}=\omega _{21}-m_{1}\Omega _{1}$ and $\Delta _{2}=\omega
_{23}-m_{2}\Omega _{2}$.

Therefore if we express the center of mass position in terms of the creation
and annihilation operators of the two-dimensional trap namely
\begin{equation}
\hat{x}=\Delta x(\hat{\psi}_{1}^{\dagger }+\hat{\psi}_{1}),\qquad \text{and}%
\qquad \hat{y}=\Delta y(\hat{\psi}_{2}^{\dagger }+\hat{\psi}_{2}).
\end{equation}%
where $\Delta x=(\hbar /2\upsilon _{1}m)^{1/2}=\eta _{1}/k_{1}$ and $\Delta
y=(\hbar /2\upsilon _{2}m)^{1/2}=\eta _{2}/k_{2}$ are the widths of the
bi-dimensional potential ground states, in the $x$ and $y$ directions ($\eta
_{i}$ is called Lamb--Dicke parameter describing the localization of the
spatial extension of the center-of-mass in $i^{\underline{th}}$ direction),
and $m$ is the mass of the ion.

Making use of the special form of Baker-Hausdorff theorem \cite{blo92} the
operator $\exp [i\eta (\hat{\psi}_{1}^{\dagger }+\hat{\psi}_{1})]$ may be
written as a product of operators i.e. $\exp (i\eta (\hat{\psi}^{\dagger }+%
\hat{\psi}))=\exp \left( \frac{\eta ^{2}}{2}[\hat{\psi}^{\dagger },\hat{\psi}%
]\right) \exp \left( i\eta \hat{\psi}^{\dagger }\right) \exp \left( i\eta
\hat{\psi}\right) .$ The physical processes implied by the various terms of
the operator
\begin{equation}
\exp \left( i\eta \left( \hat{\psi}^{\dagger }+\hat{\psi}\right) \right)
=\exp \left( \frac{-\eta ^{2}}{2}\right) \sum_{n=0}^{\infty }\frac{\left(
i\eta \right) ^{n}\hat{\psi}^{\dagger n}}{n!}\sum_{m=0}^{\infty }\frac{%
\left( i\eta \right) ^{m}\hat{\psi}^{m}}{m!}.
\end{equation}%
may be divided into three categories (i) the terms for $n>m$ correspond to
an increase in energy linked with the motional state of center of mass of
the ion by ($n-m$) quanta, (ii) the terms with $n<m$ represent destruction
of ($m-n$) quanta of energy thus reducing the amount of energy linked with
the center of mass motion and (iii) ($n=m$), represents the diagonal
contributions. When we take Lamb-Dicke limit and apply the rotating wave
approximation discarding the rapidly oscillating terms, the effective
interaction Hamiltonian (\ref{ham}) takes the form
\begin{eqnarray}
\hat{H}_{int} &=&\hbar \gamma _{1}(t)\mathcal{E}_{p}^{(1)}(\hat{\psi}%
_{1}^{\dagger }\hat{\psi}_{1})\hat{S}_{12}\hat{\psi}_{1}^{\dagger
m_{1}}+\hbar \gamma _{2}(t)\mathcal{E}_{p}^{(2)}(\hat{\psi}_{2}^{\dagger }%
\hat{\psi}_{2})\hat{S}_{23}\hat{\psi}_{2}^{\dagger m_{2}}+\hbar \gamma
_{1}^{\ast }(t)\mathcal{E}_{p}^{(1)\ast }(\hat{\psi}_{1}^{\dagger }\hat{\psi}%
_{1})\hat{S}_{21}\hat{\psi}_{1}^{m_{1}}  \notag \\
&&+\hbar \gamma _{2}^{\ast }(t)\mathcal{E}_{p}^{(2)\ast }(\hat{\psi}%
_{2}^{\dagger }\hat{\psi}_{2})\hat{S}_{32}\hat{\psi}_{2}^{m_{2}},
\end{eqnarray}%
where $\gamma _{i}(t)$ is a new coupling parameter adjusted to be time
dependent. The other contributions are rapidly oscillating with frequency $%
\nu $ and have been disregarded. Note that in the Lamb-Dicke regime only
processes with $p=0,1$ are considered, while in the general case, the
nonlinear coupling function is derived by expanding the operator-valued mode
function as
\begin{equation}
\mathcal{E}_{k}^{(j)}(\hat{\psi}_{i}^{\dagger }\hat{\psi}_{i})=-\frac{%
\epsilon _{i}}{2}\exp \left( -\frac{\eta _{i}^{2}}{2}\right)
\sum\limits_{n=0}^{\infty }\frac{(i\eta _{i})^{2n_{i}+k}}{n_{i}!(n_{i}+k)!}%
\hat{\psi}_{i}^{\dagger n}\hat{\psi}_{i}^{n}.
\end{equation}%
Since $\mathcal{E}_{k}^{(j)}(\hat{\psi}_{i}^{\dagger }\hat{\psi}_{i})$
depends only on the quantum number $\hat{\psi}_{i}^{\dagger }\hat{\psi}_{i}$%
, in the basis of its eigenstates, $\hat{\psi}_{i}^{\dagger }\hat{\psi}%
_{i}|n_{i}\rangle =n_{i}|n_{i}\rangle $, ($n=0,1,2,...$), these operators
are diagonal, with their diagonal elements $\langle n|\mathcal{E}_{k}^{(j)}(%
\hat{\psi}_{i}^{\dagger }\hat{\psi}_{i})|n\rangle $ is given by $\mathcal{E}%
_{k}^{(j)}(n_{i})=-0.5\epsilon _{i}(n_{i}+k)!)^{-1}n_{i}!L_{n_{i}}^{k}(\eta
_{i}^{2})\exp \left( -\eta _{i}^{2}/2\right) $ where $L_{n_{i}}^{k}(\eta
_{i}^{2})$ are the associated Laguerre polynomials.

In what follows we obtain a general result regarding the solution
to the time evolution operator. Now, we expand the time evolution
operator in terms of the complete set of atomic operators as
\begin{equation}
\hat{U}(t)=\exp \left( -i\int\limits_{0}^{t}\hat{H}_{int}(\tau )d\tau
\right) .  \label{ut}
\end{equation}%
In the basis of the eigenstates, $|n_{1},n_{2}\rangle $, we can find the
elements of $\hat{U}(t)$ as $\mathfrak{S}_{ii}(n_{1},n_{2},t)=\langle i|\hat{%
U}(n_{1},n_{2},t)|j\rangle ,$ where $\hat{U}(n_{1},n_{2},t)=\langle
n_{1},n_{2}|\hat{U}(t)|n_{1},n_{2}\rangle .$ Let us discuss the problem
under consideration with $\varepsilon _{i}=\varepsilon ,$ $\phi _{j}=0,$ and
the initial conditions $\mathfrak{S}_{ii}(n_{1},n_{2},0)=1$ and $\mathfrak{S}%
_{ij}(n_{1},n_{2},0)=0,$ ($i\neq j)$. Under these conditions and after
straightforward calculations, one can find an analytic time dependent
solution in the following forms
\begin{eqnarray}
\mathfrak{S}_{11}(n_{1},n_{2},t) &=&\frac{1}{\mu _{n_{1},n_{2}}^{2}}\left\{
\frac{\gamma _{1}^{2}(n_{1}+m_{1})!}{n_{1}!}\cos \left( \mu
_{n_{1},n_{2}}\int\nolimits_{0}^{t}\gamma (\tau )d\tau \right) +\frac{\gamma
_{2}^{2}(n_{2}+m_{2})!}{n_{2}!}\right\} ,  \notag \\
\mathfrak{S}_{22}(n_{1},n_{2},t) &=&\cos \left( \mu
_{n_{1},n_{2}}\int\nolimits_{0}^{t}\gamma (\tau )d\tau \right) ,  \notag \\
\mathfrak{S}_{33}(n_{1},n_{2},t) &=&\frac{1}{\mu _{n_{1},n_{2}}^{2}}\left\{
\frac{\gamma _{2}^{2}(n_{2}+m_{2})!}{n_{2}!}\cos \left( \mu
_{n_{1},n_{2}}\int\nolimits_{0}^{t}\gamma (\tau )d\tau \right) +\frac{\gamma
_{1}^{2}(n_{1}+m_{1})!}{n_{1}!}\right\} ,  \notag \\
\mathfrak{S}_{12}(n_{1},n_{2},t) &=&\frac{-i\gamma _{1}}{\mu _{n_{1},n_{2}}}%
\sqrt{\frac{(n_{1}+m_{1})!}{n_{1}!}}\sin \left( \mu
_{n_{1},n_{2}}\int\nolimits_{0}^{t}\gamma (\tau )d\tau \right) ,  \notag \\
\mathfrak{S}_{13}(n_{1},n_{2},t) &=&\frac{\gamma _{1}\gamma _{2}}{\mu
_{n_{1},n_{2}}^{2}}\left\{ \cos \left( \mu
_{n_{1},n_{2}}\int\nolimits_{0}^{t}\gamma (\tau )d\tau \right) -1\right\} ,
\notag \\
\mathfrak{S}_{23}(n_{1},n_{2},t) &=&-\frac{i\gamma _{2}^{\ast }}{\mu
_{n_{1},n_{2}}}\sqrt{\frac{(n_{2}+m_{2})!}{n_{2}!}}\sin \left( \mu
_{n_{1},n_{2}}\int\nolimits_{0}^{t}\gamma (\tau )d\tau \right) .
\label{exact}
\end{eqnarray}%
The rest of $\mathfrak{S}_{ij}(n_{1},n_{2},t)=\left( \mathfrak{S}%
_{ji}(n_{1},n_{2},t)\right) ^{\ast }$ and the generalized Rabi frequency $%
\mu _{n_{1},n_{2}}$ is given by
\begin{equation}
\mu _{n_{1},n_{2}}=\sqrt{\frac{\gamma _{1}^{2}\mathcal{E}%
_{1}^{2}(n_{1})(n_{1}+m_{1})!}{n_{1}!}+\frac{\gamma _{2}^{2}\mathcal{E}%
_{2}^{2}(n_{2})(n_{2}+m_{2})!}{n_{2}!}}.
\end{equation}%
We have thus completely determined the exact solution of a three-level
trapped ion in the presence of time-dependent modulated function. Let's
observe that putting in particular $m_{i}=1,$ $\gamma (\tau )=1$ and looking
at the present solution (\ref{exact}), we recover the time-independent
three-level system discussed in \cite{mes03,yoo85}.

The (internal level) ionic dynamics depend on the distributions of initial
excitations of both the field and the center-of-mass vibrational motion,
given by $\rho _{F}(0)$ \ and $\rho _{A}(0),$ respectively. For instance,
through a unitary evolution operator, the final state $\rho (t)$ can be
calculated in the following expression
\begin{equation}
\rho (t)=\hat{U}(t)\left( \rho _{A}(0)\otimes \rho _{F}(0)\right) U^{\dagger
}(t).
\end{equation}%
Having obtained the explicit form of the final state of the system $\rho (t)$%
, we can discuss any statistical property of the system. We will turn our
attention to the time evolution of linear entropy and entanglement, when the
field is initially in the Fock state or coherent state.

\section{Entanglement dynamics}

The entanglement can be described by the linear entropy or the von-Neumann
entropy \cite{vid00}. \ The most prominent choice of pure state entanglement
measures is the von-Neumann entropy \ \cite{vid00,pho88,pho91,pho91a}
\begin{equation*}
S\left( \rho _{A(F)}\right) =-tr\left( \rho _{A(F)}\ln \rho _{A(F)}\right) ,
\end{equation*}%
of the reduced density matrix, often simply called the entanglement $E(\psi
)=S\left( \rho _{A(F)}\right) $ of the pure state $|\psi \rangle $. We work
with the linear entropy which is convenient to calculate \ \cite{ber03},
which is given by
\begin{equation}
S_{A}\left( t\right) =1-tr_{A}\left( \rho _{A}^{2}(t)\right) ,  \label{lin}
\end{equation}%
which ranges from $0$ for a pure state to $1$ for a maximally entangled
state and $tr_{A}$ denotes the trace over the subsystem $A$. The linear
entropy is generally a simpler quantity to calculate than the von Neumann
entropy as there is no need for diagonalization and can be considered as a
very useful operational measure of the atomic state purity.
\begin{figure}[tbph]
\begin{center}
\includegraphics[width=8cm]{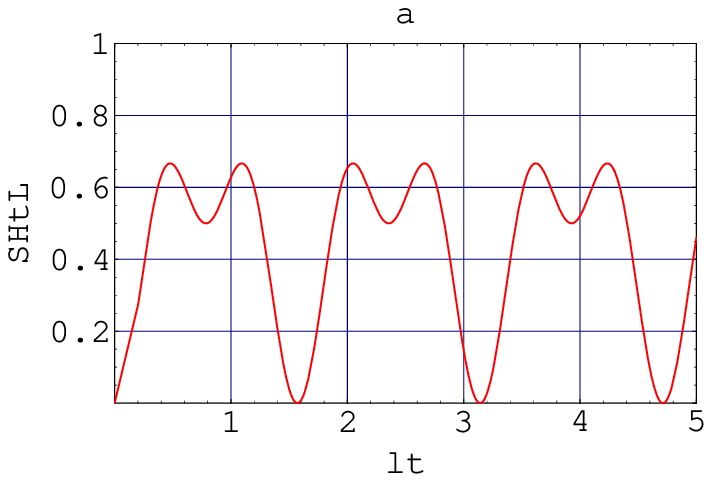} \includegraphics[width=8cm]{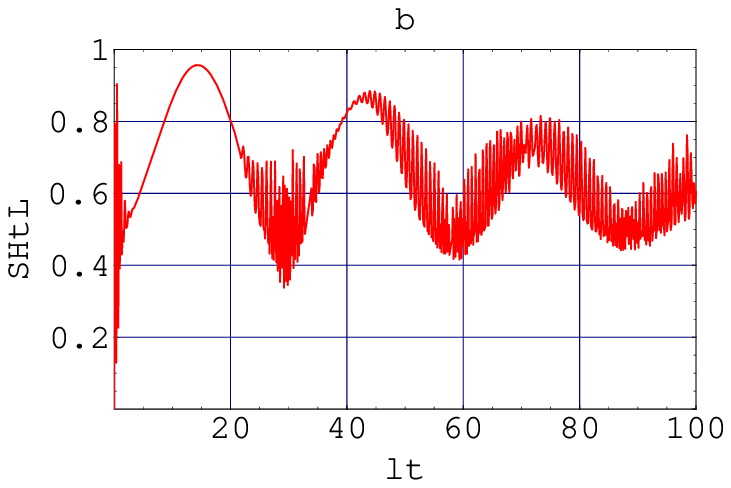}
\end{center}
\caption{Example of the numerical calculation of $S_{A}(t)$ as a function of
the scaled time $\protect\lambda t,$ ($\protect\lambda =\protect\gamma _{1}=%
\protect\gamma _{2}).$ The parameters $m_{i}=1,$ $\protect\gamma (\protect%
\tau )=1$ and different initial states of the field, where, (a) Fock state
with ($n_{i}=0)$ and (b) coherent state with ($\overline{n}_{i}=0).$}
\end{figure}

In figure 2, numerical results for the time evolution of the linear entropy
for an initial Fock state of the fields ($|n,m\rangle ,$ with $n=m=0)$ have
been presented. In the typical experiments at NIST \cite{lei00}, a single $%
^{9}Be^{+}$ ion is stored in a RF Paul trap with a secular frequency along $%
\widehat{x}$ of $\nu /2\pi \simeq 11.2$ MHz, providing a spread of the
ground state wave function of $\Delta x\simeq 7$ nm, with a Lamb-Dicke
parameter of $\eta \simeq 0.202$. The two laser beams, with $0.5$ W in each
one, are approximately detuned $\Delta /2\pi \simeq 12$ GHz, so that $\gamma
_{i}/2\pi \simeq 475$ kHz. With these data we find $\epsilon _{i}\simeq 0.01$%
, so they can be considered as small parameters. The case of the effective
vacuum is quite interesting where the linear entropy oscillates between
zeros and a maximum value, in this case $S_{A}\left( t\right) \simeq 0.65$
(see figure 2a). In fact the linear \ entropy attains the zero value (i.e.,
disentanglement) when the trapped ion is either in its upper or lower states
(i.e., pure state) while strong entanglement occurs when the inversion is
equal to zero. On other words, due to initial Fock state, the entanglement
reaches its maximum value and drops to zero periodically, which opens the
door for a possible application of the present model in constructing a
quantum logic gate.

The immediate question now is, if different initial states of the field are
considered, is the periodic behavior of the linear entropy and zeros
entanglement for such states still exist? To answer this question we make
use of a coherent state as an initial state of the field and find a general
entanglement feature, captured in equation (\ref{lin}) and illustrated in
figure 2b. Once the initial state of the field is considered to be a
coherent state the situation is changed drastically (see figure 2b). It is
obvious that the time evolution of linear entropy behaves as that of
standard single-photon Jaynes-Cummings model and oscillates irregularly with
the time. At the early times the linear entropy from zero evolves to its
local maximum value ( $\simeq 0.96$). In this process, the three
level-trapped ion and the fields are always entangled. Although increasing
the mean photon number leads to strong entanglement (maximum value of
entanglement), however the maximum value of the entanglement also varies and
occurs for some short period of time. This indicates that in a regime where
coherent state is considered, the underlying states are highly entangled.

\begin{figure}[tbph]
\begin{center}
\includegraphics[width=8cm]{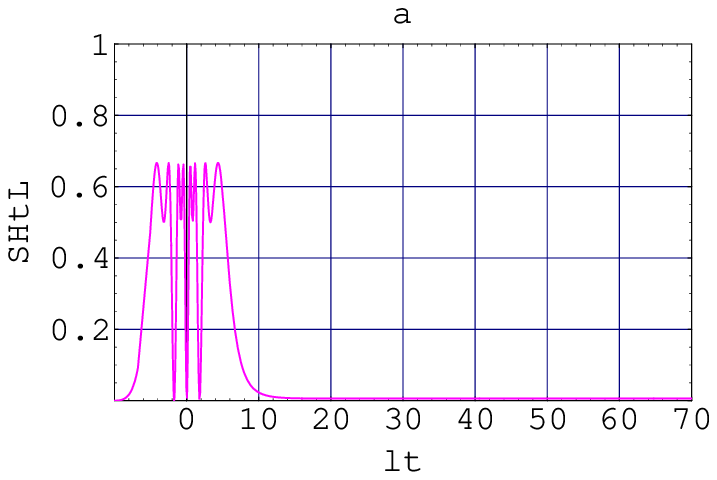} \includegraphics[width=8cm]{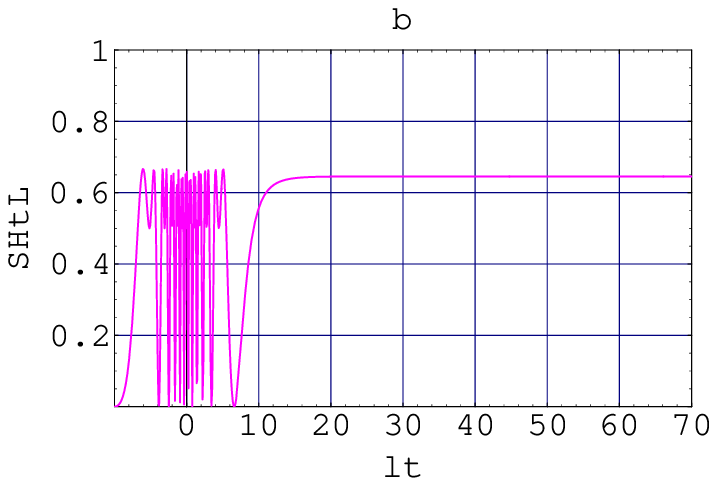}
\end{center}
\caption{The same as figure 2a but in this figure we consider the
modulated function to be time-dependent of the form
(\protect\ref{time}), the initial time is $t_{0}=-10\protect\tau $
and different values of $n_{i},$ where (a) $n_{i}=0$ and (b)
$n_{i}=15.$}
\end{figure}

Let us consider the modulated function $\gamma (t)$ to be time-dependent of
the form \cite{pra92,das99}
\begin{equation}
\gamma (t)=\sec \text{h}\left( \frac{t}{2\tau }\right) .  \label{time}
\end{equation}%
In this form the coupling increases from a very small value at large
negative times to a peak at time $t=0$, to decrease exponentially at large
times. Thus, depending on the value of $\tau $ and the initial time $t_{0}$,
various limits such as adiabatically or rapidly increasing (for $%
t_{0}<\lambda t\leq 0$) or decreasing (for $0\leq \lambda t<t_{0}$) coupling
can be conveniently studied. This allows us to investigate, analytically,
the effect of transients in various different limits of the effect of
switching the interaction on and off in the ion-field system. The vanishing
of the interaction at large positive times leads to the levelling out of the
inversion. \ It should be noted that the time dependence specified in (\ref%
{time}) is one of a class of generalized interactions that offered
analytical solutions.

In figure 3 the changes in linear entropy vs the dimensionless quantity $%
\lambda t$ is plotted when the modulated function is taken to be
time-dependent as in equation (\ref{time}). An intriguing result found in
figure 3a, where linear entropy is plotted with an initially Fock state of
the fields ($|n_{1},n_{2}\rangle ,$ with $n_{1}=n_{2}=0)$ \ with a
time-dependent modulated function, showing clearly the sudden death of
entanglement at $\lambda t\simeq 10.2$. A remarkable property of such
initial state setting is that entanglement can fall abruptly to zero for a
very long time and the entanglement will not be recovered i.e. the state
will stay in the disentanglement separable state. On the other hand, \ we
notice that the long-living entanglement can be obtained with large values
of the initial Fock state numbers, such as $|n_{1}=15,n_{2}=15\rangle ,$
(see figure 3b). Therefore, the initial Fock states placed at this point is
a suitable choice to investigate entanglement dynamics for different initial
number of photons for the fields. It's not surprising to find that the
number of oscillations is increased for higher $m_{i}$. At the period $%
-10\leq \lambda t\leq 10,$ the slight difference lies on the number of
oscillations only, while for the later times (say $\lambda t>10),$ the
situation becomes completely different.
\begin{figure}[tbph]
\begin{center}
\includegraphics[width=8cm]{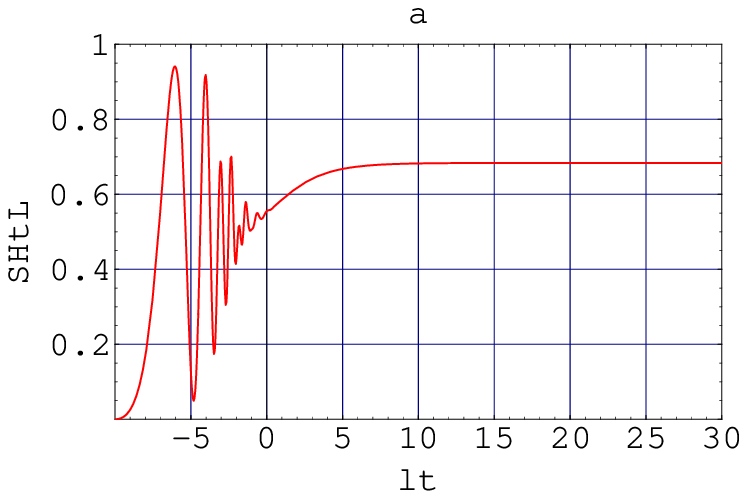} \includegraphics[width=8cm]{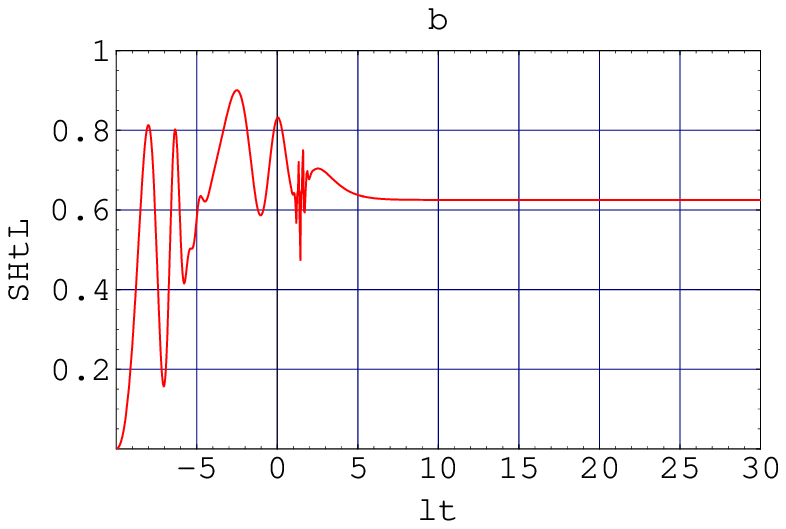}
\end{center}
\caption{ The same as figure 3 but in this figure we consider an initially
coherent state field with an average photon number of $\overline{n}_{i}=10$
and various value of the number of quanta, where (a) $m_{i}=1$ and (b) $%
m_{i}=2$.}
\end{figure}

The above results and connections are very intriguing, and lead us to ask
what is the role played by the initial state in obtaining these associated
phenomena of the entanglement. In order to answer to this question, we
consider different initial state in figure 4. This figure shows the linear
entropy with an initially coherent state field with an average photon number
of $\overline{n}_{i}=10$ for various number of quanta. The initial time is $%
t_{0}=-10\tau $, so the interaction starts at a fairly low value, peaks and
then drops off again. It is interesting to mention here that, as time goes
on long-livid entanglement is observed. As a particular but striking enough
example we have considered the same value of the system parameters which
have been considered in the time-independent case. We have analyzed the
long-lived entanglement by considering a multi-photon interaction ($%
m_{1}=m_{2}=2)$ in figure 4b. This case is similar to a situation where both
$m_{1}=m_{2}=1$, because both linear entropies in figure 4a and 4b rise and
lower together although they are not equal. The only difference between the
two cases is that the maximum long-living entanglement for one photon case
is $S(t)\simeq 0.69,$ but in the two-photon case is $S(t)\simeq 0.62.$ Also,
figure 4b demonstrates that the linear entropy peaks show a lowering of the
local maximum at the interaction period $-10\leq \lambda t\leq 0$.

All these results confirm the possibility of a practical observation of
time-dependence of the modulated function effects for creating sudden death
or long-lived entanglement. \textrm{Based on such sensitivity and some other
evidence, we suspect that the analytical results presented here, could be
attained for different configurations of the three-level systems. }

\section{Conclusions}

Summarizing, we have investigated the dynamics of quantum entanglement for a
trapped ion-laser field interaction. An explicit expression is given for a
time-independent case and compared with previous studies. Through a
three-level trapped ion system we have shown that the commonly assumed
initial state setting may affect entanglement in a very different manner.
This study reveals that the time-dependent modulated function can be used
for generating either entanglement sudden death or long-lived entanglement
depending on a proper manipulation of the initial state setting. We hope the
presented results can be useful for the ongoing theoretical and experimental
efforts in multi-levels particles interaction. Hence, despite the
considerable progresses on which we have reported here, a panoply of
challenging open questions awaits solution, what simply reflects the
decoherence effect as well as the cavity decay or atomic decay.

\textbf{Acknowledgment}

I would like to thank Prof. J. H. Eberly for his helpful comments
and suggestions.

\end{document}